\documentclass[amssymb,amsmath,onecolumn,showpacs,aps,superscriptaddress]{revtex4-1}
\usepackage{amsmath}
\usepackage{amsfonts}
\usepackage{amssymb}
\usepackage{dcolumn}
\usepackage{bm}
\usepackage[dvips]{graphicx}
\usepackage{epsfig}

\begin{document}

\title{Loschmidt echo in many-spin systems: contrasting time-scales of local
and global measurements.}

\author{Pablo R. Zangara}
\address{Instituto de F\'{i}sica Enrique Gaviola (CONICET-UNC) and Facultad
de Matem\'{a}tica, Astronom\'{i}a y F\'{i}sica, Universidad Nacional
de C\'{o}rdoba, 5000, C\'{o}rdoba, Argentina}
\author{Denise Bendersky}
\address{Instituto de F\'{i}sica Enrique Gaviola (CONICET-UNC) and Facultad
de Matem\'{a}tica, Astronom\'{i}a y F\'{i}sica, Universidad Nacional
de C\'{o}rdoba, 5000, C\'{o}rdoba, Argentina}
\author{Patricia R. Levstein}
\address{Instituto de F\'{i}sica Enrique Gaviola (CONICET-UNC) and Facultad
de Matem\'{a}tica, Astronom\'{i}a y F\'{i}sica, Universidad Nacional
de C\'{o}rdoba, 5000, C\'{o}rdoba, Argentina}
\author{Horacio M. Pastawski}
\address{Instituto de F\'{i}sica Enrique Gaviola (CONICET-UNC) and Facultad
de Matem\'{a}tica, Astronom\'{i}a y F\'{i}sica, Universidad Nacional
de C\'{o}rdoba, 5000, C\'{o}rdoba, Argentina}
\email{horacio@famaf.unc.edu.ar}



%

\begin{abstract}
A local excitation in a quantum many-spin system evolves deterministically.
A time-reversal procedure, involving the inversion of the signs of every
energy and interaction, should produce the excitation revival. This idea,
experimentally coined in NMR, embodies the concept of the Loschmidt echo
(LE). While such an implementation involves a single spin autocorrelation $%
M_{1,1}$, i.e. a local LE, theoretical efforts have focused on the study of
the recovery probability of a complete many-body state, referred here as
global or many-body LE $M_{MB}$. Here, we analyze the relation between these
magnitudes, in what concerns to their characteristic time scales and their
dependence on the number of spins $N$. We show that the global LE can be
understood, to some extent, as the simultaneous occurrence of $N$
independent local LEs, i.e. $M_{MB}\sim \left( M_{1,1}\right) ^{N/4}$. This
extensive hypothesis is exact for very short times and confirmed numerically
beyond such a regime. Furthermore, we discuss a general picture of the decay
of $M_{1,1}$ as a consequence of the interplay between the time scale that
characterizes the reversible interactions ($T_{2}$) and that of the
perturbation ($\tau _{\Sigma }$). Our analysis suggests that the short time
decay, characterized by the time scale $\tau _{\Sigma }$, is greatly
enhanced by the complex processes that occur beyond $T_{2}$ . This would
ultimately lead to the experimentally observed $T_{3},$ which was found to
be roughly independent of $\tau _{\Sigma }$ but closely tied to $T_{2}$.
\end{abstract}

\maketitle

\section{Introduction}

If an ink drop falls into a pond, the stain diffuses away until no trace of
it remains whatsoever. One may naturally say that such a process is in fact
irreversible. In the microscopic world, similar phenomena are also
ubiquitous. For instance, let us consider a many-spin quantum system in
thermal equilibrium where a local polarization excess is injected. Then,
this excitation would spread all over as consequence of spin-spin
interactions. Such an apparently irreversible process is known as \textit{%
spin diffusion} \cite{BlumeHubbard1970,forsterbook} and it can lead the
system back to equilibrium. However, this naive picture has its limitations.
On the one hand, spreading is not always the rule as there are physical
situations where the initial excitation does not vanish. This is the case of
Anderson localization \cite{altshuler2006,Zangara2015PIP} or when the
excitation remains topologically protected \cite{Wen_book}. On the other
hand, even in the cases where the system seems to have reached an
equilibrium state, the unitarity of quantum dynamics ensures a precise
memory of the non-equilibrium initial condition. Then, if some experimental
protocol could reverse the many-body dynamics, it would drive the system
back to the initial non-equilibrium state \cite{Hahn_atomicMemory}. Such a
general idea defines the Loschmidt echo (LE), which embodies the various
time-reversal procedures implemented in nuclear magnetic resonance (NMR) 
\cite{prosen,Jacquod, scholarpedia}.

The first NMR time-reversal experiment was performed by Hahn in the 1950's 
\cite{Hahn1950}. The procedure, known as spin echo, reverses the precession
dynamics of each independent spin around its local magnetic field by
inverting the sign of the Zeeman energy. However, the\ sign of the energy
associated to the spin-spin interactions is not inverted and, accordingly,
the echo signal is degraded. Such a decay occurs within the time scale $%
T_{2} $ that characterizes the spin-spin interactions. Indeed, these
interactions determine the survival of a spin excitation at short times as $%
\sim 1-\left( t/T_{2}\right) ^{2}$ and its later complex dynamics generating
a diffusive spreading. By the early 1970's, Kessemeier, Rhim, Pines and
Waugh implemented the reversal of the dynamics induced by the spin-spin
dipolar interaction \cite{Kessemeier1971,Rhim1971}. This results in the
\textquotedblleft magic echo\textquotedblright\, which indicates the
recovery of a global polarization state. Two decades later, Ernst and
collaborators introduced the \textquotedblleft polarization
echo\textquotedblright\ \cite{Ernst1992}. There, a local excitation injected
in a many-spin system is let to evolve, then time-reversed and finally
detected locally at the initial spot. While the success of these time
reversal echoes unambiguously evidenced the deterministic nature of
spin-dynamics in NMR, it is clear that the reversal is unavoidably degraded
by uncontrolled internal or environmental degrees of freedom or by
imperfections in the pulse sequences. Furthermore, the degradation seems to
occur in a time scale, say $T_{3}$, much shorter than a naive estimation of
the characteristic scale of these perturbations, say $\tau _{\Sigma }$.
Then, the question that arises is whether the complexity inherent to a large
number of correlated spins would enhance the fragility of the procedure
under perturbations.

A next generation of experiments in organic crystals \cite%
{patricia98,usaj-physicaA,MolPhys} seemed to confirm that the experimental $%
T_{3}$ never exceeds a few times $T_{2}$. In other words, $T_{3}$ keeps tied
to the time scale that characterizes the reversible many-body interaction.
This led to postulate that in an infinite many-spin system the complex
dynamics could favor the action of any small non-inverted interaction that
perturbs the reversal procedure. Thus, reversible interactions become
determinant for the irreversibility rate. This constitutes our \textit{%
Central Hypothesis of Irreversibility}. Such a wisdom is further reinforced
by the natural association of many-body complexity with a form of chaos \cite%
{Flambaum2000,Flambaum2001d} and\ the confirmation that quantum dynamics of
classically chaotic systems should manifest a dynamical instability \cite%
{peres1984} which leads to an \textit{environment-independent decoherence
rate}\cite{jalpa,Cookprb2004}.

During the last decade, solid-state NMR kept providing a versatile testing
bench to study time reversal in large spin arrays \cite%
{Suter2004,Boutis2012,Cappellaro2013,Claudia2014,Alvarez2014experimental}.
As a matter of fact, a standard experiment involves a crystalline sample
with an infinitely large number of spins. In contrast, the numerical test of
many spin dynamics has to be restricted to strictly finite systems \cite%
{Zangara2012,Fine2014}. While this appears to be a major limitation, it
allows the analysis of a situation that the experiments cannot achieve:
moving progressively from small systems to larger ones with a fully
controlled perturbation. The expectancy is that a sort of finite size
scaling may allow to identify an emergent mechanism that rules reversibility
in the thermodynamic limit \cite{Zangara2015PRA}. As in the experiments, the
witness for such a transition should be the LE as measured by a single spin
autocorrelation function $M_{1,1}$, i.e. a local polarization. For short we
call $M_{1,1}$\ the \textit{local LE}. It is not difficult to probe that $%
\Pi _{1,1}\equiv (M_{1,1}+1)/2$ is the probability that a given spin, say
the $1^{\mathrm{st}}$, remains up after the whole procedure. Besides, in a
case of a spin excitation in a 1D chain with $XY$ interactions \cite%
{mesoECO-PRL1995,mesoECO-exp,ERNST-CPL-1997} $M_{1,1}$ precisely coincides 
\cite{Zangara2012} with the global overlap of two one-body wave functions as
defined for semiclassical systems \cite%
{jalpa,Jacquod-Beenakker-2001,Jacquod2002}. The square of the overlap
between the initial and final many-body wave functions, $M_{MB}$, defines a 
\textit{global} or \textit{many-body LE}. It is important to notice that $%
M_{MB}$ has not been addressed experimentally, but nevertheless it is a
natural magnitude in theory \cite{prosen2007_rev,Santos2014a,Santos2014b}.
Thus, we are left without a precise relation between the object of
theoretical studies and experimental ones, i.e. $M_{MB}$\ and $M_{1,1}$\
respectively. This missing link is the central question we address in this
paper.

Here, we consider a system of $N$ spins whose initial state is given by a
local excitation injected in an high temperature state. Firstly, we discuss
the formal relation between $M_{MB}$\ and $M_{1,1}$, which is derived
exactly at least for very short times. In particular, we assess how the $N$%
-dependence or extensivity of $M_{MB}$ is evidenced in the time scales
involved. This leads us to hint that the revival of a many-spin state
results from the recovery of each individual spin configuration, much as if
they were statistically independent events. Since in the initial high
temperature state there are $N/2$ spins up, their rough statistical
independence would lead to a behavior of the sort of $M_{MB}\sim \left( \Pi
_{1,1}\right) ^{N/2}\sim \left( M_{1,1}\right) ^{N/4}$. This is confirmed by
the numerical evaluation of the LE in a specific spin model.

Furthermore, we discuss a general picture beyond the short-time regime,
where the decay of $M_{1,1}$ results from the interplay between the time
scale that characterizes the reversible interactions ($T_{2}$) and that of
the perturbation ($\tau _{\Sigma }$). This would ultimately lead to the
experimentally observed $T_{3}$. In such a sense, our analysis provides a
conceptual hinge between the theoretical and the experimental realms.

The paper is organized as follows. In Sec. \ref{Sec_formulation} we
introduce the LE framework:\ the initial state and the time-reversal
procedure. Here, we define both the local and global LE. In Sec. \ref%
{Sec_locglob} we compute the short time expansions for the local LE and its
non-local contributions (in particular, the many-body LE). This allows us to
discuss a general picture of the LE decay in terms of the times scales that
characterize $M_{MB}$\ and $M_{1,1}$. The dependence with $N$ is discussed
in terms of the extensivity of $M_{MB} $ and statistical independence of the
local autocorrelations. In Sec. \ref{Sec_1Dmodel} we assess our expectancies
by a numerical evaluation of the LE in a spin system. Section \ref%
{Sec_conclu} summarizes our main conclusions and some of the important open
questions in the field.

\section{The Loschmidt echo in spin systems.\label{Sec_formulation}}

Let us first specify the initial condition of a \textquotedblleft local
excitation in many-spin system\textquotedblright . We consider $N$ spins $%
1/2 $ in an infinite temperature state, i.e. completely depolarized mixture,
plus a locally injected polarization,

\begin{equation}
\hat{\rho}_{0}=\frac{1}{2^{N}}(\mathbf{\hat{I}}+2\hat{S}_{1}^{z}).
\label{rho_inicial}
\end{equation}%
Here, the spin $1$ is polarized while the others are not, i.e. $tr[\hat{S}%
_{i}^{z}\hat{\rho}_{0}]=\frac{1}{2}\delta _{i,1}$. Such an initial state can
be experimentally implemented not only in NMR \cite{Cappellaro2014_review}
but also in cold atoms \cite{Bloch2013}.

\begin{figure}[!h]
\centering\includegraphics[width=5.0in]{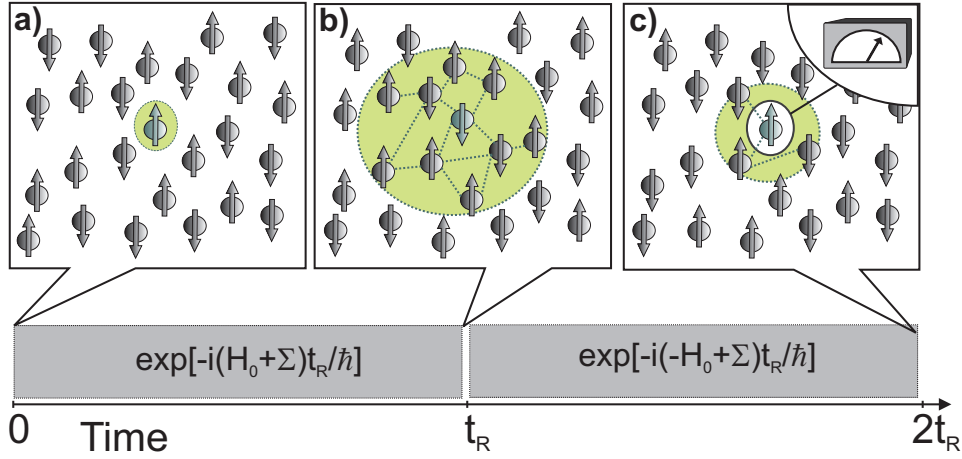} 
\caption{The scheme of the local LE, $M_{1,1}(t)$. (a) The initial state is
given by a local excitation in a high temperature spin system, as stated in
equation (\protect\ref{rho_inicial}). The system is let to evolve under the
Hamiltonian $\hat{H}_{0}+\hat{\Sigma}$ and the excitation diffuses all over
until a time $t=t_{R}$ (b). At that time, a time reversal is performed,
leading to a backward evolution ruled by $-\hat{H}_{0}+\hat{\Sigma}$. At
time $t=2t_{R}$ (c), a local measurement is performed at the initial spot.}
\label{Fig_LE_standard}
\end{figure}

As in the early LE experiments \cite{patricia98,usaj-physicaA,MolPhys}, our
numerical evaluation focuses on an imperfect time reversed evolution of the
excitation, followed by a local measurement. The procedure is depicted in
Fig. \ref{Fig_LE_standard}. A many-spin Hamiltonian $\hat{H}_{0}$ rules the 
\textit{forward} evolution of the system up to a certain time $t_{R}$. At
that moment, an inversion of the sign of $\hat{H}_{0}$ is performed, leading
to a symmetric \textit{backward }evolution. Nevertheless, there are
unavoidable perturbations, denoted by $\hat{\Sigma}$, that could arise from
the incomplete control of the Hamiltonian, acting on both periods. Thus,
evolution operators for these $t_{R}$-periods are $\hat{U}%
_{+}^{{}}(t_{R})=\exp [-\frac{\mathrm{i}}{\hbar }(\hat{H}_{0}+\hat{\Sigma}%
)t_{R}]$ and $\hat{U}_{-}^{{}}(t_{R})=\exp [-\frac{\mathrm{i}}{\hbar }(-\hat{%
H}_{0}+\hat{\Sigma})t_{R}]$ respectively. It is quite practical to define
the LE\ operator as:

\begin{equation}
\hat{U}_{LE}^{{}}(2t_{R})=\hat{U}_{-}^{{}}(t_{R})\hat{U}_{+}^{{}}(t_{R}),
\label{Ule}
\end{equation}%
which produces an imperfect refocusing at time $2t_{R}$. A local measurement
of the polarization, performed at site $1$, defines the local LE:

\begin{equation}
M_{1,1}(t)=2tr[\hat{S}_{1}^{z}\hat{U}_{LE}^{{}}(t)\hat{\rho}_{0}\hat{U}%
_{LE}^{\dag }(t)]=2tr[\hat{S}_{1}^{z}\hat{\rho}_{t}].
\label{autocorrelacion1}
\end{equation}%
Here, we choose as free variable $t=2t_{R}$, the total elapsed time in the
presence of the perturbation. The time dependence of $\hat{\rho}_{t}$ in the
Schr\"{o}dinger picture is,

\begin{equation}
\hat{\rho}_{t}=\hat{U}_{LE}^{{}}(t)\hat{\rho}_{0}\hat{U}_{LE}^{\dag }(t).
\label{autocorrelacion11}
\end{equation}


Using equation (\ref{rho_inicial}), and after some algebraic manipulation, the LE
can be explicitly written as a correlation function:

\begin{equation}
M_{1,1}(t)=\frac{1}{2^{N-2}}tr[\hat{U}_{LE}^{\dag }(t)\hat{S}_{1}^{z}(0)\hat{%
U}_{LE}^{{}}(t)\hat{S}_{1}^{z}(0)]=\frac{tr[\hat{S}_{1}^{z}(t)\hat{S}%
_{1}^{z}(0)]}{tr[\hat{S}_{1}^{z}(0)\hat{S}_{1}^{z}(0)]}.
\label{autocorrelacion2}
\end{equation}%
Here, the time dependence is written according to the Heisenberg picture,

\begin{equation}
\hat{S}_{1}^{z}(t)=\hat{U}_{LE}^{\dag }(t)\hat{S}_{1}^{z}(0)\hat{U}%
_{LE}^{{}}(t).  \label{autocorrelacion21}
\end{equation}%
Notice that equation (\ref{autocorrelacion2}) is an explicit correlation function
at the same site but different times, i.e. an \textit{auto}correlation. This
kind of a correlation has been recently employed to address localization
phenomena in spin systems \cite{Zangara2013PRB,knap2015,Zangara2015PIP} and
it generalizes the standard one employed to assess spin diffusion \cite%
{LevPasCalvo1989}. In terms of the Hilbert--Schmidt inner product between
the initial and the time evolved density matrices, i.e. equations (\ref%
{rho_inicial}) and (\ref{autocorrelacion11}) respectively, the LE can be
written as \cite{usaj-physicaA,PazZurek2003,Bendersky2013}:

\begin{equation}
M_{1,1}(t)=2^{N}tr[\hat{\rho}_{0}\hat{\rho}_{t}]-1=2\frac{tr[\hat{\rho}_{0}%
\hat{\rho}_{t}]}{tr[\hat{\rho}_{0}\hat{\rho}_{0}]}-1,
\label{autocorrelacion3}
\end{equation}%
which, in the present case, progressively decays from $1$ to $0$ as it
occurs with the statistical overlap between two wave packets in the standard
LE definition \cite{jalpa}.

Equivalent expressions for the LE can be derived decomposing the statistical
state in a simpler basis. In order to proceed with the pure state
decomposition of $\hat{\rho}_{0}$, we consider the computational Ising basis 
$\left\{ \left\vert \beta _{i}\right\rangle \right\} $, also known as $S^{z}$%
-decoupled basis. Additionally, we define the set $\mathcal{A}$ of indexes $j
$ that label basis states which have the $1^{\mathrm{st}}$ spin pointing up,
i.e. $j\in \mathcal{A}$ $\Leftrightarrow \hat{S}_{1}^{z}\left\vert \beta
_{j}\right\rangle =+\frac{1}{2}\left\vert \beta _{j}\right\rangle $. It is
straightforward to verify that $\hat{\rho}_{0}=\sum_{j\in \mathcal{A}%
}2^{-(N-1)}\left\vert \beta _{j}\right\rangle \left\langle \beta
_{j}\right\vert $. Then, as introduced in Ref. \cite{mesoECO-PRL1995},

\begin{equation}
M_{1,1}(t)=2\left[ \sum_{i\in \mathcal{A}}\sum_{j\in \mathcal{A}}\frac{1}{%
2^{N-1}}\left\vert \left\langle \beta _{j}\right\vert \hat{U}%
_{LE}^{{}}(t)\left\vert \beta _{i}\right\rangle \right\vert ^{2}-\frac{1}{2}%
\right] =2\left[ \Pi _{1,1}(t)-\frac{1}{2}\right] .  \label{autoco2}
\end{equation}%
Here, $\Pi _{1,1}(t)$ denotes the \textit{probability} that the $1^{st}$
spin keeps pointing up after a time $t$. After some manipulation,

\begin{equation}
\begin{split}
M_{1,1}(t)& =2\left[ \sum_{i\in \mathcal{A}}\sum_{j\in \mathcal{A}}\frac{1}{%
2^{N-1}}\left\vert \left\langle \beta _{j}\right\vert \hat{U}%
_{LE}^{{}}(t)\left\vert \beta _{i}\right\rangle \right\vert ^{2}-\frac{1}{2}%
\right] \\
& =\left[ \sum_{i\in \mathcal{A}}\frac{1}{2^{N-1}}\left( \vphantom{\sum_{i%
\in \mathcal{A}}}\left\vert \left\langle \beta _{i}\right\vert \hat{U}%
_{LE}^{{}}(t)\left\vert \beta _{i}\right\rangle \right\vert ^{2}+\right.
\right. \\
& \ \left. \left. \qquad \qquad \qquad \quad +\sum_{j\in \mathcal{A}\text{ (}%
j\neq i)}\left\vert \left\langle \beta _{j}\right\vert \hat{U}%
_{LE}^{{}}(t)\left\vert \beta _{i}\right\rangle \right\vert ^{2}-\sum_{j\in 
\mathcal{B}}\left\vert \left\langle \beta _{j}\right\vert \hat{U}%
_{LE}^{{}}(t)\left\vert \beta _{i}\right\rangle \right\vert ^{2}\right) %
\right] .
\end{split}
\label{contribuciones}
\end{equation}%
Here, $\mathcal{B}$ stands for the complement of $\mathcal{A}$, i.e. $j\in 
\mathcal{B}$ $\Leftrightarrow \hat{S}_{1}^{z}\left\vert \beta
_{j}\right\rangle =-\frac{1}{2}\left\vert \beta _{j}\right\rangle $. One can
naturally identify and define the two terms that contribute to the local
polarization $M_{1,1}(t)$. The first sum in equation (\ref{contribuciones})
stands for the average probability of revival of the many-body states,
denoted by $M_{MB}(t)$,

\begin{equation}
M_{MB}(t)=\sum_{i\in \mathcal{A}}\frac{1}{2^{N-1}}\left\vert \left\langle
\beta _{i}\right\vert \hat{U}_{LE}^{{}}(t)\left\vert \beta _{i}\right\rangle
\right\vert ^{2}.  \label{autocorrelacion51}
\end{equation}%
The second sum in equation (\ref{contribuciones}) represents the average
probability of changing the configuration of any spin except the $1^{\mathrm{%
st}}.$ The third sum stands for the average probability that the $1^{\mathrm{%
st}}$ spin has actually flipped, i.e. of all those processes that do not
contribute to $M_{1,1}(t)$. Then, the processes that contribute to $%
M_{1,1}(t)$ but not to $M_{MB}(t)$ are denoted as:

\begin{equation}
M_{X}(t)=\sum_{i\in \mathcal{A}}\frac{1}{2^{N-1}}\left( \sum_{j\in \mathcal{A%
}\text{ (}j\neq i)}\left\vert \left\langle \beta _{j}\right\vert \hat{U}%
_{LE}^{{}}(t)\left\vert \beta _{i}\right\rangle \right\vert ^{2}-\sum_{j\in 
\mathcal{B}}\left\vert \left\langle \beta _{j}\right\vert \hat{U}%
_{LE}^{{}}(t)\left\vert \beta _{i}\right\rangle \right\vert ^{2}\right).
\label{autocorrelacion52}
\end{equation}%
This balance of probabilities leads to the appropriate asymptotic behavior
of $M_{1,1}(t)$ according to the symmetries that constrain the evolution.
The identification

\begin{equation}
M_{1,1}(t)=M_{MB}(t)+M_{X}(t)  \label{LE_contributions}
\end{equation}%
is a crucial step for the following discussions.

If we use the identity $\hat{S}_{1}^{z}=\hat{S}_{1}^{+}\hat{S}_{1}^{-}-\frac{%
1}{2}\mathbf{\hat{I}}$ in equation (\ref{autocorrelacion2}), the invariance of
the trace under cyclic permutations ensures that $tr[\hat{S}_{1}^{z}(t)\hat{S%
}_{1}^{z}(0)]=tr[\hat{S}_{1}^{-}(0)\hat{S}_{1}^{z}(t)\hat{S}_{1}^{+}(0)]-%
\frac{1}{2}tr[\hat{S}_{1}^{z}(t)]$. Since $tr[\hat{S}_{1}^{z}(t)]=tr[\hat{S}%
_{1}^{z}(0)]=0$, then:

\begin{eqnarray}
M_{1,1}(t) &=&2\sum_{i}\frac{1}{2^{N-1}}\left\langle \beta _{i}\right\vert 
\hat{S}_{1}^{-}(0)\hat{U}_{LE}^{\dag }(t)\hat{S}_{1}^{z}(0)\hat{U}%
_{LE}^{{}}(t)\hat{S}_{1}^{+}(0)\left\vert \beta _{i}\right\rangle  \notag \\
&=&2\sum_{i\in \mathcal{A}}\frac{1}{2^{N-1}}\left\langle \beta
_{i}\right\vert \hat{U}_{LE}^{\dag }(t)\hat{S}_{1}^{z}\hat{U}%
_{LE}^{{}}(t)\left\vert \beta _{i}\right\rangle ,
\label{M11_promedioEnsamble}
\end{eqnarray}%
which is indeed an explicit way to rewrite equation (\ref{autocorrelacion1}) in
the form of an ensemble average. Remarkably, since $\hat{S}_{1}^{z}$ is a
local (\textquotedblleft one-body\textquotedblright ) operator, its
evaluation in equation (\ref{M11_promedioEnsamble}) can be replaced by the
expectation value in a single superposition state \cite{Alv-parallelism},

\begin{equation}
M_{1,1}(t)=2\left\langle \Psi _{neq}\right\vert \hat{U}_{LE}^{\dag }(t)\hat{S%
}_{1}^{z}\hat{U}_{LE}^{{}}(t)\left\vert \Psi _{neq}\right\rangle ,
\label{autocorrelacion7}
\end{equation}%
where: 
\begin{equation}
\left\vert \Psi _{neq}\right\rangle =\sum\limits_{i\in \mathcal{A}}\frac{1}{%
\sqrt{2^{N-1}}}e^{\mathrm{i}\varphi _{i}}\text{\ }\left\vert \beta
_{i}\right\rangle .  \label{neqsup}
\end{equation}%
Here, $\varphi _{i}^{{}}$ is a random phase uniformly distributed in $%
[0,2\pi )$. As a matter of fact, the state defined in equation (\ref{neqsup}) is
a random superposition that can successfully mimic the dynamics of ensemble
calculations and provides a quadratic speedup of computational efforts \cite%
{Alv-parallelism,Fine2013,Pineda2014}.

\section{The Loschmidt echo dynamics\label{Sec_locglob}}

\subsection{Short time expansions and beyond.}

In order to analyze the $N$-dependence of the LE and its time scales, we
compute here the short time expansion of the magnitudes $M_{1,1}(t)$, $%
M_{MB}(t)$ and $M_{X}(t)$. Up to $2^{nd}$ order in time,

\begin{eqnarray}
M_{1,1}(t=2t_{R}) &=&2\sum_{i\in \mathcal{A}}\frac{1}{2^{N-1}}\left\langle
\beta _{i}\right\vert \hat{U}_{LE}^{\dag }(t)\hat{S}_{1}^{z}\hat{U}%
_{LE}^{{}}(t)\left\vert \beta _{i}\right\rangle  \notag \\
&=&1-\left( t/\hbar \right) ^{2}\sum_{i\in \mathcal{A}}\frac{1}{2^{N-1}}%
\left( \left\langle \beta _{i}\right\vert \hat{\Sigma}^{2}\left\vert \beta
_{i}\right\rangle -2\left\langle \beta _{i}\right\vert \hat{\Sigma}\hat{S}%
_{1}^{z}\hat{\Sigma}\left\vert \beta _{i}\right\rangle \right) +\mathcal{O}%
\left( \left( t/\hbar \right) ^{3}\right) .  \label{chplg_st1}
\end{eqnarray}%
Similarly, the leading contributions to $M_{MB}(t)$ and $M_{X}(t)$ are:

\begin{eqnarray}
M_{MB}(t) &=&\sum_{i\in \mathcal{A}}\frac{1}{2^{N-1}}\left\vert \left\langle
\beta _{i}\right\vert \hat{U}_{LE}^{{}}(t)\left\vert \beta _{i}\right\rangle
\right\vert ^{2}  \notag \\
&=&1-\left( t/\hbar \right) ^{2}\sum_{i\in \mathcal{A}}\frac{1}{2^{N-1}}%
\left( \left\langle \beta _{i}\right\vert \hat{\Sigma}^{2}\left\vert \beta
_{i}\right\rangle -\left\langle \beta _{i}\right\vert \hat{\Sigma}\left\vert
\beta _{i}\right\rangle ^{2}\right) +\mathcal{O}\left( \left( t/\hbar
\right) ^{3}\right) ,  \label{chplg_st2}
\end{eqnarray}%
and%
\begin{eqnarray}
M_{X}(t) &=&\sum_{i\in \mathcal{A}}\frac{1}{2^{N-1}}\left( \sum_{j\in 
\mathcal{A}\text{ (}j\neq i)}\left\vert \left\langle \beta _{j}\right\vert 
\hat{U}_{LE}^{{}}(t)\left\vert \beta _{i}\right\rangle \right\vert
^{2}-\sum_{j\in \mathcal{B}}\left\vert \left\langle \beta _{j}\right\vert 
\hat{U}_{LE}^{{}}(t)\left\vert \beta _{i}\right\rangle \right\vert
^{2}\right)  \notag \\
&=&\left( t/\hbar \right) ^{2}\sum_{i\in \mathcal{A}}\frac{1}{2^{N-1}}\left(
2\left\langle \beta _{i}\right\vert \hat{\Sigma}\hat{S}_{1}^{z}\hat{\Sigma}%
\left\vert \beta _{i}\right\rangle -\left\langle \beta _{i}\right\vert \hat{%
\Sigma}\left\vert \beta _{i}\right\rangle ^{2}\right) +\mathcal{O}\left(
\left( t/\hbar \right) ^{3}\right) .  \label{chplg_st3}
\end{eqnarray}

Let us consider a generic \textit{secular }(i.e. polarization conserving)
perturbation $\hat{\Sigma}$ given by a Hamiltonian with an arbitrary
anisotropy $\alpha $,%
\begin{equation}
\hat{\Sigma}=\sum_{i,j}^{N}(J_{\Sigma})_{ij}\left[ 2\alpha \hat{S}_{i}^{z}%
\hat{S}_{j}^{z}-\left( \hat{S}_{i}^{x}\hat{S}_{j}^{x}+\hat{S}_{i}^{y}\hat{S}%
_{j}^{y}\right) \right] .  \label{chplg_pert}
\end{equation}%
This is still quite general since even a double quantum perturbation ($\hat{S%
}_{i}^{+}\hat{S}_{j}^{+}+\hat{S}_{i}^{-}\hat{S}_{j}^{-}$, which does not
conserve polarization) can be reduced to a secular one by the truncating
effects of radiofrequency fields \cite{Zangara2015PRA}. In addition, we do
not consider here the case $\left[ \hat{\Sigma},\hat{S}_{1}^{z}\right] =0$
(e.g. pure Ising or on-site diagonal disorder), since in such a condition
the first non-trivial order in time is the $4^{th}$ (see Appendix). Then,
the following identities hold:%
\begin{eqnarray}
\sum_{i\in \mathcal{A}}\frac{1}{2^{N-1}}\left\langle \beta _{i}\right\vert 
\hat{\Sigma}^{2}\left\vert \beta _{i}\right\rangle &=&2N\sigma ^{2}\left( 
\frac{\alpha ^{2}}{4}+\frac{1}{8}\right) ,  \label{chplg_st4} \\
\sum_{i\in \mathcal{A}}\frac{1}{2^{N-1}}\left\langle \beta _{i}\right\vert 
\hat{\Sigma}\hat{S}_{1}^{z}\hat{\Sigma}\left\vert \beta _{i}\right\rangle
&=&2N\sigma ^{2}\left( \frac{\alpha ^{2}}{8}+\frac{1}{16}\right) -\frac{1}{2}%
\sigma ^{2},  \label{chplg_st5} \\
\sum_{i\in \mathcal{A}}\frac{1}{2^{N-1}}\left\langle \beta _{i}\right\vert 
\hat{\Sigma}\left\vert \beta _{i}\right\rangle ^{2} &=&2N\sigma ^{2}\frac{%
\alpha ^{2}}{4}.  \label{chplg_st6}
\end{eqnarray}%
Here, $\sigma ^{2}$ stands for the average local second moment of $\hat{%
\Sigma}$,%
\begin{equation}
\sigma _{{}}^{2}=\frac{1}{N}\sum_{i=1}^{N}\sigma _{i}^{2}=\frac{1}{N}%
\sum_{i=1}^{N}\left[ \sum_{j(\neq i)}^{N}\left( \frac{(J_{\Sigma})_{ij}}{2}%
\right) ^{2}\right] .  \label{chplg_sigmalocal}
\end{equation}%
Being a \textit{perturbation}, we notice that $\sigma ^{2}$ is much smaller
than the average local second moment $\sigma _{0}^{2}$ of the unperturbed
Hamiltonian $\hat{H}_{0}$. In terms of time scales, 
\begin{equation}
T_{2}=\hbar /\sqrt{\sigma _{0}^{2}}\ll \hbar /\sqrt{\sigma _{{}}^{2}}=\tau
_{\Sigma }.  \label{time-comparison}
\end{equation}%
The identities in equations (\ref{chplg_st4}), (\ref{chplg_st5}) and (\ref%
{chplg_st6}) lead to

\begin{equation}
M_{1,1}(t)=1-\left( t/\tau _{\Sigma }\right) ^{2}+\mathcal{O}\left( \left(
t/\hbar \right) ^{3}\right) ,  \label{chplg_st7}
\end{equation}%
and,

\begin{eqnarray}
M_{MB}(t) &=&1-\frac{1}{4}N\left( t/\tau _{\Sigma }\right) ^{2}+\mathcal{O}%
\left( \left( t/\hbar \right) ^{3}\right) ,  \label{chplg_st8} \\
M_{X}(t) &=&\left( \frac{N-4}{4}\right) \left( t/\tau _{\Sigma }\right) ^{2}+%
\mathcal{O}\left( \left( t/\hbar \right) ^{3}\right) .  \label{chplg_st9}
\end{eqnarray}%
These expansions hold for $t<\left( \tau _{\Sigma }/N\right) $. Beyond such
a very short time regime, a general term in the expansion of $M_{1,1}(t)$
will be of the form

\begin{equation}
c_{(N,n)}t^{n}/\left( \tau _{\Sigma }^{k}T_{2}^{n-k}\right)
\label{termino_gral}
\end{equation}%
with $k\geq 2$ and the coefficient $c_{(N,n)}$ described by combinatorial
numbers of increasing size that depend on the topology of the interactions
(e.g. see \cite{Claudia2014,Zamar2015}). Since the experimental set up
corresponds to the limit described by equation (\ref{time-comparison}), this
expansion will be dominated by terms with the lowest possible order in the
weak interaction, i.e. $k=2$:%
\begin{equation}
(t/\tau _{\Sigma })^{2}\left[ 1+\sum_{n}c_{(N,n)}(t/T_{2}^{{}})^{n-2}\right]
.  \label{orden_dominante}
\end{equation}%
Equation (\ref{orden_dominante}) indicates that beyond the very short time
expansion, i.e. $(\tau _{\Sigma }/N)<t<\tau _{\Sigma }$, the dependence on $%
\tau _{\Sigma }^{{}}$ becomes superseded by the diverging terms in the scale 
$T_{2}^{{}}$. This could lead to the new time scale $T_{3}$ which was seen
experimentally to be tied to $T_{2}$ as

\begin{equation}
T_{2}\lesssim T_{3}\ll \tau _{\Sigma }.  \label{comparison2}
\end{equation}%
In that sense, $T_{3}$ becomes characteristic of the complexity or
\textquotedblleft chaos\textquotedblright\ of the many spin system that
amplifies the small effect of the perturbation. In addition, it is important
to stress that, being an experimental fact, equation (\ref{comparison2})
corresponds to a system composed by infinitely many interacting spins. In
other words, equation (\ref{comparison2}) stands for the relations of time scales
in the thermodynamic limit. Quite on the contrary, any numerical simulation
involves a finite, very small indeed, number of spins where the
irreversibility rate $T_{3}$ would be essentially given by $\tau _{\Sigma}$%
. Then, the LE decay rate evaluated in a finite system would ultimately be
perturbation-dependent \cite{Fine2014}. Thus, our \textit{Central Hypothesis
of Irreversibility} would mean that equation (\ref{comparison2}) is an \textit{%
emergent} property. It should rely on the thermodynamic limit, which\
implies taking the limit $N\rightarrow \infty $ first, and then $\tau_{\Sigma}\rightarrow \infty $. The non-uniformity of these limits plays a crucial role to yield quantum phase transitions, as discussed in the context
of Anderson localization \cite{AndersonRMP-1978,Rodrigues1986,hmp-physB}.

\begin{figure}[h]
\centering\includegraphics[width=4.5in]{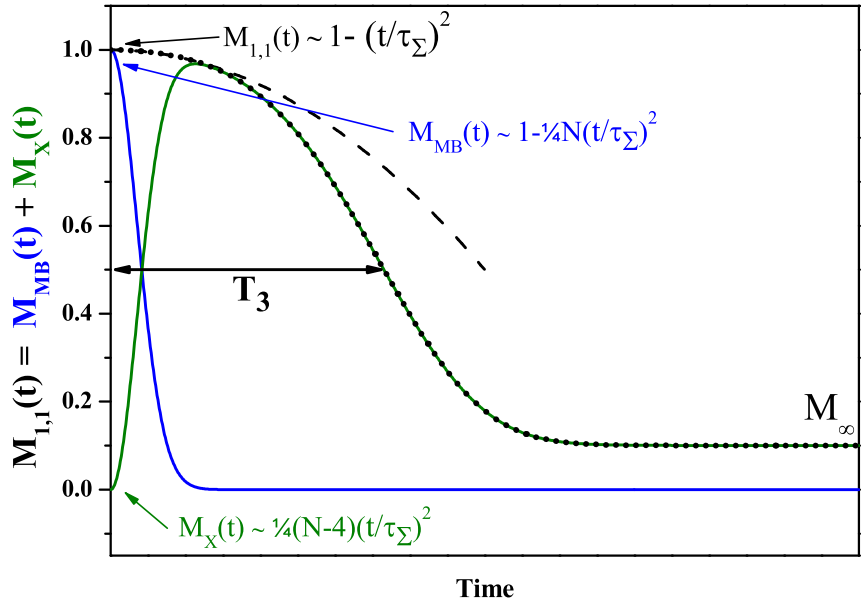} 
\caption{A pictorial scheme of the time dependence of $M_{1,1}(t)$ (dotted line) and its contributions, i.e. $M_{MB}(t)$ and $M_{X}(t)$ (solid lines). Their short-time expansions, as stated in equations  (\protect\ref{chplg_st7}), (\protect\ref{chplg_st8}) and (\protect\ref{chplg_st9}), are indicated with arrows. In particular, the expansion corresponding to the short-time behavior of $M_{1,1}(t)$ is plotted with a dashed line. (Online version in colour.)}
\label{dynamicalScheme}
\end{figure}

The physical picture described above is schematically represented in Fig. %
\ref{dynamicalScheme}. There, we show the expected interplay between $%
M_{MB}(t)$ and $M_{X}(t)$ leading to $M_{1,1}(t)$. Indeed, as stated in
equations (\ref{chplg_st8}) and (\ref{chplg_st9}), the very short time
dependence of both contributions is extensive in $N$: $M_{MB}(t)$ decreases
as $1-N\sigma _{{}}^{2}t^{2}/4$ and $M_{X}(t)$ increases as $(N-4)\sigma
_{{}}^{2}t^{2}/4$. Such a precise balance provides for the short time decay
of $M_{1,1}(t)$ given by equation (\ref{chplg_st7}), i.e. $1-\sigma
_{{}}^{2}t^{2} $. Notice that there is no reason to assume that the decay of 
$M_{MB}(t)$ would remain ruled by $\tau _{\Sigma }$. Beyond the very short
times, we expect that the time scale $T_{3}$ should also show up as $%
T_{3}/N^{\nu }$ with $\nu \sim 1$ in the decay of $M_{MB}(t)$(see below).
Furthermore, while $M_{MB}(t)$ goes monotonically to zero, $M_{X}(t)$
displays a highly non-trivial behavior. Indeed, $M_{X}(t)$ first increases
by feeding from the decay of $M_{MB}(t)$ until it reaches a maximum. This
growth indicates a progressive divergence of long-range correlations.
Thereafter, $M_{X}(t)$ should decay accounting for the fact that the state
remains properly normalized. This is precisely what $M_{1,1}(t)$ measures: a
conserved polarization that ultimately distributes uniformly within the spin
system. In an isolated finite system this implies the asymptotic plateau $%
M_{\infty }\sim 1/N$. As pointed above, the decay of both $M_{1,1}(t)$ and $%
M_{X}(t)$ occurs in a time scale $T_{3}$, which according to equation (\ref%
{orden_dominante}), is somewhat longer but close to the \textquotedblleft
diffusion\textquotedblright\ time $T_{2}$. This is the regime captured
experimentally.

\subsection{The extensive decay hypothesis\label{Sec_scaling}}

The previous short-time expansions provide a hint on the scaling relation
between the local LE, $M_{1,1}(t),$ and the global one as embodied by $%
M_{MB}(t)$. In particular, let us first compare the probability of
refocusing the configuration (up or down) of a single spin, i.e. $\Pi
_{1,1}(t)$, and the probability of refocusing a complete many-spin state $%
M_{MB}(t)$. If the refocusing of each individual spin could be treated as an
independent event, then the scaling between $\Pi _{1,1}$ and $M_{MB}$ would
be extensive in $N$,

\begin{equation}
\left( \Pi _{1,1}(t)\right) ^{N/2}\simeq M_{MB}(t).
\label{scaling_locglobal}
\end{equation}%
Here, the factor $1/2$ in the exponent comes from equation (\ref{rho_inicial}),
i.e. the initial high temperature state, where basically half of the spins
point up, and half of them point down.\textbf{\ }Then, one can resort to the
picture of a lattice gas where\textbf{\ }$N/2$ particles jump among $N$
lattice sites. As in the well known Jordan-Wigner transformation \cite%
{Lieb-Mattis1961}, a fermion is associated to a spin pointing up and a
vacancy corresponds to a spin pointing down. Thus, the microstate of the gas
is completely described by the position of $N/2$ particles.

Strictly speaking, the notion of \textit{extensiveness} corresponds to standard thermodynamic quantities such as the entropy of the system. In addition, as discussed in \cite{MolPhys}, $S=-ln(M_{1,1}(t))$ is precisely a measure of the entropy. Then, the validity of equation (\ref{scaling_locglobal}) implies an extensivity relation between the entropy per spin and the total entropy of the system. 

According to equations (\ref{autoco2}) and (\ref{chplg_st7}),

\begin{equation}
\Pi _{1,1}(t)=1-\frac{1}{2}\left( t/\tau _{\Sigma }\right) ^{2}+\mathcal{O}%
\left( \left( t/\hbar \right) ^{3}\right) ,  \label{scaling_aux}
\end{equation}%
which in turn, up to $2^{\mathtt{nd}}$ order in time, implies $\Pi _{1,1}(t)\simeq
\left( M_{1,1}(t)\right) ^{1/2}$. Thus, equation (\ref{scaling_locglobal}) yields

\begin{equation}
\left( M_{1,1}(t)\right) ^{N/4}\simeq M_{MB}(t).  \label{scaling_locglobal2}
\end{equation}%
This is precisely the relation verified between equations (\ref{chplg_st7}) and (%
\ref{chplg_st8}).

One might expect that beyond the very short-time decay, individual spin
autocorrelations deviate from the statistical independence. However, this
deviation will still have a local nature and therefore the $N$-extensivity
would remain valid. Indeed, we propose%
\begin{equation}
\left( M_{1,1}(t)\right) ^{\eta }\simeq M_{MB}(t),
\label{scaling_locglobal3}
\end{equation}%
where the exponent $\eta $ would be some appropriate function $\eta =\eta
(N,t)$. Our \textquotedblleft extensive decay hypothesis\textquotedblright\
implies that $\eta $ factorizes:%
\begin{equation}
\eta (N,t)=N\times f(t),  \label{factoriza_exponente}
\end{equation}%
where $f(t)$ stands for a function that encloses information of the
correlations originated by the system dynamics. Additionally,

\begin{equation}
\lim_{t\rightarrow 0^{+}}f(t)=\frac{1}{4}.  \label{limite_ft}
\end{equation}%
is required in order to recover equation (\ref{scaling_locglobal2}), i.e. the
statistical independence.

\section{A 1D model.\label{Sec_1Dmodel}}

The physical picture described above is discussed here under the light of a
specific model. In particular, we assess the validity of equations (\ref%
{scaling_locglobal2}) and (\ref{factoriza_exponente}). We consider a 1-D
spin chain with an anisotropic interaction described by:

\begin{equation}
\hat{H}_{0}=\sum_{i=1}^{N-1}J_{0}\left( \frac{1}{2}\hat{S}_{i}^{z}\hat{S}%
_{i+1}^{z}+\hat{S}_{i}^{x}\hat{S}_{i+1}^{x}+\hat{S}_{i}^{y}\hat{S}%
_{i+1}^{y}\right) ,  \label{chplg_H1vecinos}
\end{equation}%
with periodic boundary conditions, i.e. a ring configuration. Here, $J_{0}$
stands for the natural units of the spin-spin interaction energy. As a
perturbation $\hat{\Sigma}$ we choose a next nearest neighbors interaction
described by:

\begin{equation}
\hat{\Sigma}=\sum_{i=1}^{N-2}J_{\Sigma}\left( \frac{1}{2}\hat{S}_{i}^{z}\hat{%
S}_{i+2}^{z}+\hat{S}_{i}^{x}\hat{S}_{i+2}^{x}+\hat{S}_{i}^{y}\hat{S}%
_{i+2}^{y}\right) .  \label{chplg_H2vecinos}
\end{equation}
Such a perturbation appears naturally when one attempts to build an
effective one-body dynamics from linear crystals with dipolar interactions 
\cite{elenaMQC}. This is also the case in a regular crystal, when the
natural non-secular dipole-dipole terms are truncated by the Zeeman energy
of the radiofrequency irradiation, which ultimately leads to effective
secular two-body next nearest neighbors interactions \cite{Zangara2015PRA}.

The local second moments $\sigma ^{2}$ and $\sigma _{0}^{2}$ of $\hat{\Sigma}
$ and $\hat{H}_{0}$ respectively can be evaluated as in equation (\ref%
{chplg_sigmalocal}):

\begin{eqnarray}
\sigma ^{2} &=&\frac{1}{2}\left( J_{\Sigma} \right) ^{2},
\label{chplg_sigmaModelo} \\
\sigma _{0}^{2} &=&\frac{1}{2}\left( J_{0}^{{}}\right) ^{2},
\label{chplg_sigmaH0}
\end{eqnarray}%
and constitute the main energy scales of our problem.

\begin{figure}[h]
\centering\includegraphics[width=3.8in]{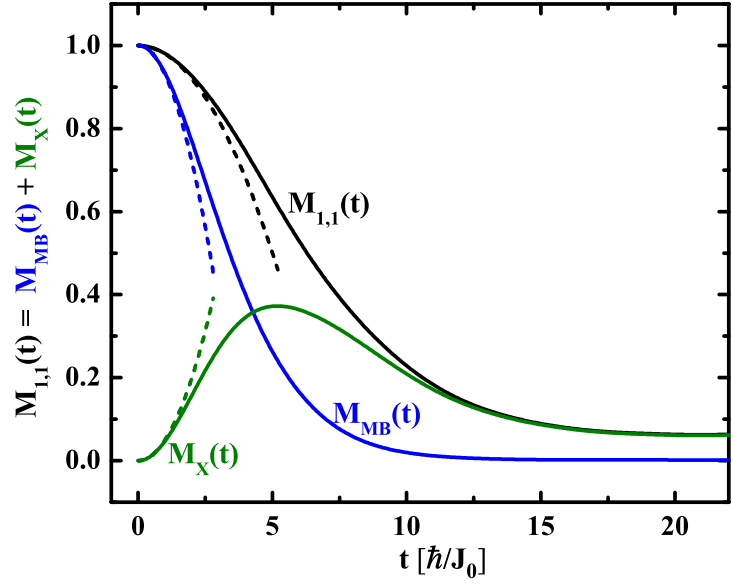} 
\caption{The local LE and its non local contributions. $M_{1,1}(t)$, $M_{MB}(t)$ and $M_{X}(t)$ correspond to the solid lines as indicated by the labels in the figure. $N=14$, $J_{\Sigma }=0.1J_{0}^{}$. The short-time expansions given in equations (\protect\ref{chplg_st7}), (\protect\ref{chplg_st8}) and (\protect\ref{chplg_st9}) are shown in dashed lines. (Online version in colour.)}
\label{Ecos_dinamica}
\end{figure}

In Fig. \ref{Ecos_dinamica} we plot $M_{1,1}(t)$, $M_{MB}(t)$ and $M_{X}(t)$
for the particular choice $J_{\Sigma }=0.1J_{0}^{{}}$. Short-time expansions
given in equations (\ref{chplg_st7}), (\ref{chplg_st8}) and (\ref{chplg_st9})
are evaluated according to equation (\ref{chplg_sigmaModelo}). It is observed
that $M_{MB}(t)$ vanishes for long times.\ Actually, a close observation
shows that $M_{MB}(t\rightarrow \infty )\sim \mathcal{O}(2^{-N})$ (data not
shown). In addition, notice that $M_{X}(t\rightarrow \infty )\sim 1/N$. Such
an asymptotic contribution provides for the equidistribution of the spin
polarization $M_{1,1}(t\rightarrow \infty )\sim 1/N$. This long-time
saturation corresponds to the equilibration of a finite system.

In contrast to our schematic plot in Fig. \ref{dynamicalScheme}, here $%
M_{X}(t)$ does not get too close to $1$ and $M_{MB}(t)$ does not decay much
faster than $M_{1,1}(t)$. Since $M_{X}(t)$ provides for the whole $%
M_{1,1}(t) $ once that $M_{MB}(t)$ has fully decayed, the contribution of $%
M_{X}(t)$ is considerable only at long times. These effects are a
consequence of the relatively small size of the system considered. Indeed,
the case in Fig. \ref{Ecos_dinamica} corresponds to $N=14$ spins, and thus
the exponent that relates $M_{1,1}(t)$ and $M_{MB}(t)$ is quite small $%
(N/4)=3.5$. The need for larger systems indicates that revealing the
dominant orders in equation (\ref{orden_dominante}) is a major numerical
challenge that may go beyond the state-of-the-art techniques \cite%
{Dente2013CPC}.

\begin{figure}[h]
\centering\includegraphics[trim=0 0 0 0 ,clip,width=4.5in]{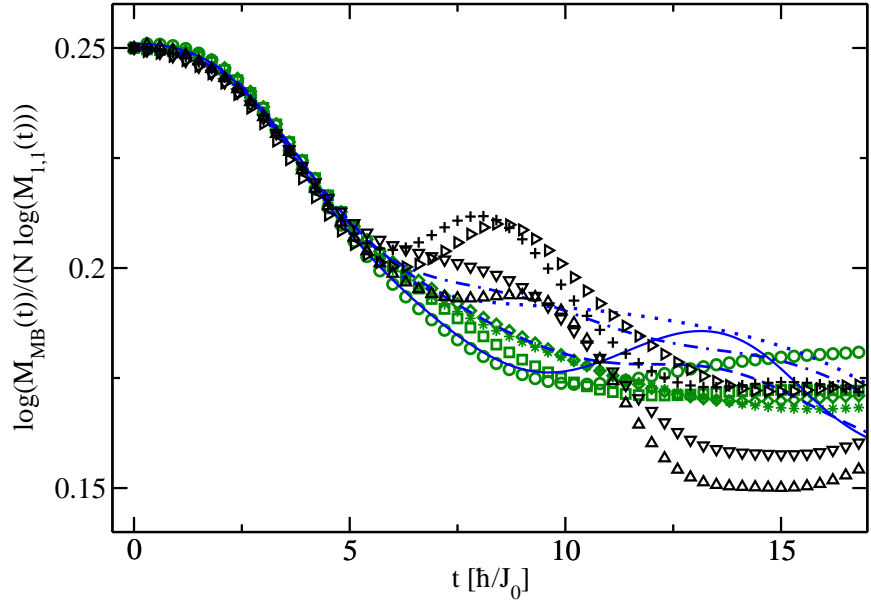} 
\caption{The relation $\log (M_{MB}(t))/\left( N\log (M_{1,1}(t))\right)$ as a function of time. For $J_{\Sigma }=0.1J_{0}^{}$ the sizes plotted are: $N=10$ (circles), $N=12$ (squares), $N=14$ (diamonds), and $N=16$ (stars). For $J_{\Sigma }=0.2J_{0}^{}$ the sizes plotted are: $N=10$ (solid line), $N=12$ (dashed line), $N=14$ (dash-dot line), and $N=16$ (dotted line). For $J_{\Sigma }=0.3J_{0}^{}$ the sizes plotted are: $N=10$ (up triangles), $N=12$ (down triangles), $N=14$ (plus signs), and $N=16$ (right triangles). (Online version in colour.) }
\label{Fig_factorizaexp}
\end{figure}

In order to assess the accuracy of the \textquotedblleft extensive decay
hypothesis\textquotedblright , in Fig. \ref{Fig_factorizaexp} we address the
scaling relation between $M_{1,1}(t)$ and $M_{MB}(t)$ discussed in Sec. \ref%
{Sec_scaling}. In particular, we try out the factorization stated in equation (%
\ref{factoriza_exponente}). By plotting $\log (M_{MB}(t))/\left( \log
(M_{1,1}(t))N\right) $ as a function of time, we observe a unique function
which does not depend on $N$ or $J_{\Sigma }$, but it has a weak dependence
on time. Such a unique curve is indeed $f(t)$ as defined in equation (\ref%
{factoriza_exponente}). This means that the extensivity relation between $%
M_{1,1}(t)$ and $M_{MB}(t)$ is confirmed. The statistical independence, in
turn, fails progressively once $f(t)$ departs from the $1/4$ factor of the
ideal relation in equations (\ref{scaling_locglobal2}) and (\ref{limite_ft}).
Since beyond the short-time regime $f(t)$ decreases with time, we conclude
that the recovery of a single spin is tied to the recovery of its neighbors.
Thus, the spins are positively correlated and the revival probability of the
complete $N$-spin state is enhanced. This argument is particularly relevant
in 1D systems.

After the onset of the saturation regime, where $M_{1,1}\sim 1/N$ and $%
M_{MB}\sim \mathcal{O}(2^{-N})$, the universal scaling naturally becomes
noisy and the curves for different $N$ and $J_{\Sigma }$ separate each
other. Since the decay is faster for larger perturbations, the appearance of
such a spurious behavior is observed to occur first for the largest value of 
$J_{\Sigma }$ considered ($J_{\Sigma }=0.3J_{0}^{{}}$, plus signs and triangles).

\section{Conclusion \label{Sec_conclu}}

We presented a detailed analysis of the LE in interacting spin systems. As
in the NMR experiments, a local version of the LE, $M_{1,1}$, is defined as
a single spin autocorrelation function. Simultaneously, we define a global
LE, $M_{MB}$, as the average of the square overlap between many-body wave
functions that evolved under perturbed Hamiltonians. While the former
constitutes a specific experimental observable, the latter has only been
assessed theoretically. Here, we showed the formal relation between both
magnitudes, as far as their characteristic time scales and $N$-dependence
are concerned.

By analyzing a short-time expansion of $M_{1,1}$and $M_{MB}$ we derived a
precise the relation between their time scales. In this regime, the decay of 
$M_{1,1}$ is given by the average local second moment of the perturbation ($%
\hbar /\tau _{\Sigma }=\sqrt{\sigma _{{}}^{2}}$), and the decay of $M_{MB}$\
by $N$ times the local scale ($N\hbar /\tau _{\Sigma }.$). This relation
hints a scaling law $M_{MB}\sim \left( M_{1,1}\right) ^{N/4}$ that accounts
for the extensivity of $M_{MB}$. In such a case, the recovery of a many-spin
state results from the recovery of each individual spin, much as if they
were independent events. The numerical evaluation in a specific spin model
shows that the exponent slightly diminishes with time, starting from the
initial $N/4$. This means that the recovery of a single spin is positively
correlated with the probability of recovery of its neighbors, and thus it
improves the probability of the revival of the complete $N$-spin state. A
precise control of these correlations may hint an experimental access to the
global autocorrelation, i.e. $M_{MB}$, just by measuring a single spin
(local) autocorrelation $M_{1,1}$. This would require an experimental
protocol capable to encode a local excitation into a correlated many-spin
state.

In addition, we discussed a general dynamical picture beyond the very
short-time regime. There, the decay of $M_{1,1}$ results from the interplay
between the time scale that characterizes the reversible interactions ($%
T_{2} $) and that of the perturbation ($\tau _{\Sigma }$). This would
ultimately lead to the experimentally observed $T_{3}$, which was found to
be roughly independent of $\tau _{\Sigma }$ but closely related to $T_{2}$.
The theoretical quest for the emergent $T_{3}$ time scale remains open and
it may be out of the reach of current numerical approaches. Assessing a fair
estimate analytically would require a detailed account of the higher order
processes that dress the quadratic term in the perturbative expansion.

Notice that our discussion lead us to identify $T_{3}$, and hence the
spin-spin interaction time $T_{2}$, as the time scales characterizing the
complexity or many-spin chaos. As such, they show up not only in the decay
of $M_{1,1}$ and $M_{MB}$, but also in the growth of $M_{X}=M_{1,1}-M_{MB}.$
Indeed, in the field of AdS/CFT there is an increasing interest in
characterizing the role of chaos in quantum dynamics \cite%
{Stanford_butterfly,Susskind2008,Susskind2015,Kitaev_charla}. There, chaos
manifests in the growth of four-body correlation functions, following an
early suggestion by Larkin and Ovchinnikov \cite{larkin1969}. They employed
semiclassical arguments to address disordered superconductors and probed
that the square dispersion of momentum should grow exponentially in a time
scale determined by the collisions with impurities, i.e. with the
unperturbed Hamiltonian (in our physical picture, $T_{2}$). Similarly, our
average multispin correlation $M_{X}$ would ultimately diverge within a time
scale $T_{3}/N$, i.e. independent of the perturbation. This is indeed a
measure of the decoherence, and hence of irreversibility, induced by
many-spin chaos. Of course, we do not have a precise characterization of
this time scale or the specific mathematical dependence on time. Thus, this
is a puzzling issue to explore in the field of many-body chaos. Besides the
obvious relevance for statistical mechanics and experimental physics, this
might also contribute to a possible pathway between quantum mechanics and
gravity.

\section*{Acknowledgments}
This work benefited from discussions with A.D. Dente and F. Pastawski. HMP
greatly acknowledges hospitality of A. Kitaev at Caltech, P.A. Lee at MIT
and V. Oganesyan at CUNY, where the issues discussed in this paper acquired
certain maturity. PRZ acknowledges M.C. Ba\~{n}uls and J.I. Cirac for their
kind hospitality at MPQ in Garching. We acknowledge financial support from CONICET, ANPCyT, SeCyT-UNC and
MinCyT-Cor. This work used computational resources from CCAD -- Universidad
Nacional de C\'{o}rdoba (http://ccad.unc.edu.ar/), in particular the
Mendieta Cluster, which is also part of SNCAD -- MinCyT, Rep\'{u}blica
Argentina.

\section{Appendix}

It is worthy to mention that very short-time expansions in equations (\ref%
{chplg_st7}), (\ref{chplg_st8}) and (\ref{chplg_st9}) do not depend on the
anisotropy $\alpha $ of the perturbation. In general, it can be proved that
if $\left[ \hat{\Sigma},\hat{S}_{1}^{z}\right] =0$ then:

\begin{equation}
M_{1,1}(t)=1-\frac{\left( t/\hbar \right)^{4}}{2^{N+3}}\sum_{i\in \mathcal{A}%
}\left( 2\left\langle \beta _{i}\right\vert \left[ \hat{\Sigma},\hat{H}_{0}%
\right] \hat{S}_{1}^{z}\left[ \hat{\Sigma},\hat{H}_{0}\right] \left\vert
\beta _{i}\right\rangle -\left\langle \beta _{i}\right\vert \left[ \hat{%
\Sigma},\hat{H}_{0}\right] ^{2}\left\vert \beta _{i}\right\rangle \right)+%
\mathcal{O}\left( \left( t/\hbar \right) ^{5}\right).  \label{a15}
\end{equation}%
This is precisely the case of a perturbation $\hat{\Sigma}$ enclosing
Anderson disorder and Ising interactions \cite{Zangara2013PRB} or
interactions with a fluctuating field \cite{LucasF}.


\end{document}